\documentstyle[prl,aps,preprint,tighten,floats,epsf,rotate]{revtex}

\newbox\rotbox

\begin{document}
\draft


\preprint{\vbox{
                                        \null\hfill\rm TRI-PP-95-86\\
                                        \null\hfill\rm nucl-th/9511021 }}
%
\title{Recovering Relativistic Nuclear Phenomenology
       from the Quark-Meson Coupling Model}
\author{Xuemin Jin and B.K.~Jennings}
\address{TRIUMF, 4004 Wesbrook Mall, \\Vancouver,
British Columbia, Canada V6T 2A3\\}
%
%
\maketitle

\begin{abstract}
  The quark-meson coupling (QMC) model for nuclear matter, which describes
  nuclear matter as non-overlapping MIT bags bound by the self-consistent
  exchange of scalar and vector mesons is modified by the introduction of a
  density dependent bag constant. The density dependence of the bag constant 
  is related to that of the in-medium effective nucleon mass through a scaling 
  ansatz suggested by partial chiral symmetry restoration in nuclear matter.
  This modification overcomes drawbacks of the QMC model and leads to the 
  recovery of the essential features of relativistic nuclear phenomenology. 
  This suggests that the modification of the bag constant in the nuclear medium 
  may play an important role in low- and medium-energy nuclear physics.
\end{abstract}
%

\newpage
\narrowtext

Since quantum chromodynamics (QCD) is believed to be the correct theory
underlying strong interactions, the physics of nuclei is, in essence, an
exercise in applied QCD.  Building connections between observed nuclear
phenomena and the interactions and symmetries of the underlying quark and gluon
degrees of freedom has become one of the principal goals of nuclear theorists.
However, due to the complexity of the low-energy QCD, knowledge of QCD has had
very little impact, to date, on the study of low- and medium-energy nuclear
phenomena.

A reasonable consensus is that the relevant degrees of freedom for nuclear
physics at low energy scales are hadrons instead of quarks and gluons. One
general approach, relativistic nuclear phenomenology which has gained much
credibility during last twenty years, is to treat nucleons in nuclear
environment as point-like Dirac particles interacting with large canceling
scalar and vector potentials. This approach has been successful in describing
the spin-observables of nucleon-nucleus scattering in the context of
relativistic optical potentials\cite{hama90,wallace87}.  Moreover, such
potentials can be derived from the relativistic impulse
approximation\cite{wallace87}. The relativistic field-theoretical models based
on nucleons and mesons, QHD, also feature Dirac nucleons interacting through
the exchange of scalar and vector mesons\cite{serot86}. QHD, at the mean-field
level, has proven to be a powerful tool for describing the bulk properties of
nuclear matter and spin-orbit splittings of finite nuclei\cite{serot86}. It is
known that the large and canceling scalar and vector potentials are central to
the success of the relativistic nuclear phenomenology.  Recent progress in
understanding the origin of these large potentials for propagating nucleons in
nuclear matter has been made via the analysis of the finite-density QCD sum
rules\cite{cohen95}.

Given the wide success of the Dirac approach in describing various low- and
medium-energy nuclear phenomena and the support from the finite-density QCD sum
rules, it is a challenge to study the relevance of the quark structure of the
nucleon to the dynamics of normal nuclei.  A few years ago,
Guichon\cite{guichon88} proposed a quark-meson coupling (QMC) model to
investigate the direct ``quark effects'' in nuclei. In this model, nuclear
matter consists of non-overlapping MIT bags interacting through the
self-consistent exchange of mesons in the mean-field approximation, and the
mesons are directly coupled to the quarks. This simple QMC model has been
refined by including nucleon Fermi motion and the center-of-mass corrections to
the bag energy \cite{fleck90} applied to variety of
problems\cite{saito94,saito92,saito94a,saito95,saito95a,song95,guichon95}.
(There have been several works that also discuss the quark effects in nuclei,
based on other effective models for the nucleon\cite{banerjee92}).

Although it provides a simple and attractive framework to incorporate the quark
structure of the nucleon in the study of nuclear phenomena, the QMC model has a
serious short-coming.  It predicts much smaller scalar and vector potentials
for the nucleon than obtained in relativistic nuclear phenomenology and
finite-density QCD sum rules. Unless there is a large isoscalar anomalous
coupling (ruled out by other considerations) this implies a much smaller
nucleon spin-orbit force in finite nuclei.  To lowest order in the nucleon
velocity and potential depth the nucleon spin-orbit potential can be obtained
in a model independent way from the strengths of the scalar and vector
potentials. The spin-orbit potential from the QMC model is too weak to
successfully explain spin-orbit splittings in finite nuclei and the
spin-observables in nucleon-nucleus scattering.

We observe that the bag constant is held to be at its free space value in the
QMC model for nuclear matter. This assumption can be questioned.  In the MIT
bag model, the bag constant denotes the vacuum energy (relative to the
perturbative vacuum), which contributes $\sim 300$ MeV to the nucleon energy
and provides the necessary pressure to confine the quarks. Thus, the bag
constant is an inseparable ingredient of the bag picture of a nucleon. When a
nucleon bag is put into the nuclear medium, the bag as a whole reacts to the
environment. As a result, the bag constant may be modified.

There is little doubt that at sufficiently high densities, the bag constant is
eventually melted away and quarks and gluons become the appropriate degrees of
freedom.  It, thus, seems reasonable that the bag constant be modified and
decrease as density increases. Moreover, it is argued in Ref.~\cite{adami93}
that the MIT bag constant is related to the energy associated with the chiral
symmetry restoration (the vacuum energy difference between the
chiral-symmetry-restored vacuum inside the bag and the broken phase outside).
Since chiral symmetry is partially restored in nuclear medium\cite{brown95},
the in-medium bag constant should drop relative to its free space
value\cite{adami93,brown95}.  This physics has been ignored in the QMC model.

In this Letter, we shall introduce the in-medium modification of the bag 
constant in the QMC model for nuclear matter. We relate the in-medium bag
constant to the in-medium effective nucleon mass through a scaling ansatz
suggested by partial chiral symmetry restoration in nuclear matter.
We find that the essential features of the relativistic nuclear phenomenology, 
in particular the large canceling scalar and vector potentials and hence strong 
spin-orbit force for the nucleon, can be recovered when the decrease of the bag constant 
in medium is taken into account.  This suggests that the drop of the bag constant 
in nuclear medium relative to its free space value may play an important role in low- and
medium-energy nuclear physics.

In the QMC model, the nucleon in nuclear medium is assumed to be a static 
spherical MIT bag in which quarks interact with the scalar and vector fields,
$\overline{\sigma}$ and $\overline{\omega}$, and these fields are treated as
classical fields in the mean field approximation. (Here we only consider up and down
quarks.)  The quark field, $\psi_q(x)$, inside the bag then satisfies the
equation of motion:
\begin{equation}
\left[i\,\rlap{/}\partial-(m_q^0-g_\sigma^q\, \overline{\sigma})
-g_\omega^q\, \overline{\omega}\,\gamma^0\right]\,\psi_q(x)=0\ ,
\label{eq-motion}
\end{equation}
where $m_q^0$ is the current quark mass, and $g_\sigma^q$
and $g_\omega^q$ denote the quark-meson coupling constants.
The normalized ground state for a quark in the bag is given
by\cite{guichon88,fleck90,saito94}
\begin{equation}
\psi_q(t,{\bf r})={\cal N}\, e^{-i\epsilon_q t/R}
\left(
\begin{array}{c}
j_0(xr/R)
\\*[7.2pt]
i\,\beta_q\, {\bf \sigma\cdot \hat{r}}\, j_1(xr/R)
\end{array}
\right)\, {\chi_q\over \sqrt{4\pi}}\ ,
\label{wave-function}
\end{equation}
where 
$\epsilon_q=\Omega_q + g_\omega^q\, \overline{\omega}\, R$
and $\beta_q=\sqrt{(\Omega_q-R\, m_q^*)/
(\Omega_q\, +R\, m_q^*)}$,
with $\Omega_q\equiv \sqrt{x^2+(R\, m_q^*)^2}$, 
$m_q^*=m_q^0-g_\sigma^q\, \overline{\sigma}$, $R$ the bag radius, 
and  $\chi_q$ the quark spinor. The normalization factor is given by
${\cal N}^{-2}=2\, R^3 \, j_0^2(x)\left[
\Omega_q(\Omega_q-1)+R\, m_q^*/2\right]/x^2$.
The $x$ value is determined by the 
boundary condition at the bag surface, $j_0(x)=\beta_q\, j_1(x)$.
%
%

The energy of a static bag consisting of three ground state quarks 
can be expressed as
\begin{equation}
E_{\rm bag}=3\, {\Omega_q\over R}-{Z\over R}
+{4\over 3}\,  \pi \, R^3\,  B\ ,
\label{ebag}
\end{equation}
where $Z$ is a parameter which accounts for zero-point motion
and $B$ is the bag constant. In the calculations to follow, we 
use $R_0$, $B_0$ and $Z_0$ to denote the corresponding bag parameters
for the free nucleon. After the corrections of spurious 
c.m. motion in the bag, the effective mass of a nucleon bag at rest
is taken to be\cite{fleck90,saito94}
\begin{equation}
M_N^*=\sqrt{E_{\rm bag}^2-\langle p_{\rm cm}^2\rangle}\ ,
\label{eff-mn}
\end{equation}
where $\langle p_{\rm cm}^2\rangle=\sum_q \langle p_q^2\rangle$ and
$\langle p_q^2\rangle$ is the expectation value of the quark momentum
squared, $(x/R)^2$. 

The QMC model assumes that both $Z$ and $B$ are independent of 
density. (The bag radius is determined by the equilibrium condition
for the bag, see below.) This assumption is unjustified. As argued in 
Ref.~\cite{adami93}, the MIT bag constant is essentially the energy
associated with the chiral symmetry restoration. Therefore, one expects 
the bag constant to drop relative to its free space value as a consequence 
of partial chiral symmetry restoration in the nuclear medium. Of course, 
one has to invoke model descriptions in order to obtain a quantitative 
estimate for the reduction of the bag constant. According to the scaling ansatz
advocated by Brown and Rho\cite{brown91}, the bag constant should scale like
$B/B_0\simeq \Phi^4$ \cite{adami93,brown95}. Here $\Phi$ denotes the 
universal scaling, and $\Phi\sim m^*_\rho/m_\rho\simeq f^*_\pi/f_\pi\simeq 
(M^*_N/M_N)^{2/3}$ has been suggested in Ref.~\cite{brown95}. Motivated by these
suggestions, we introduce the following scaling ansatz
\begin{equation}
{B\over B_0}=\left({M_N^*\over M_N}\right)^{\kappa/3}\ ,
\label{ansatz}
\end{equation}
for the in-medium bag constant. The case $\kappa=0$ corresponds to 
the usual QMC model. Note that combining Eqs.~(\ref{ebag}), (\ref{eff-mn}), and 
(\ref{ansatz}) yields a self-consistency condition for $B$.
In principle, the parameter $Z$ may also be modified in the nuclear medium.
However, unlike the bag constant, it is unclear how $Z$ changes with
the density as $Z$ is not directly related to chiral symmetry. Here 
we assume that the medium modification of $Z$ is small at low and
moderate densities and take $Z=Z_0$.

The bag radius is determined by the equilibrium condition for the bag, 
$\partial M_N^*/\partial R = 0$.
In free space, one may fix $M_N$ at 
its experimental value $939$ MeV and use the equilibrium condition
to determine the bag parameters. For several choice of bag radius,
$R_0 = 0.6, 0.8, 1.0$ fm, the results for $B_0^{1/4}$ and $Z_0$
are $188.1, 157.5, 136.3$ MeV and $2.030,1.628,1.153$, respectively.

The total energy per nucleon at finite density, $\rho_N$, can be 
written as\cite{saito94}
\begin{equation}
E_{\rm tot} = {\gamma\over (2\pi)^3\, \rho_N} 
\int^{k_F}\, d^3 k \sqrt{M_N^{* 2}+{\bf k}^2}
+{g_\omega^2\over 2 m_\omega^2}\,\rho_N
+{m_\sigma^2\over 2\,\rho_N}\overline{\sigma}^2\ ,
\label{etot}
\end{equation}
where $\gamma$ is the spin-isospin degeneracy, and $\gamma=4$ for
symmetric nuclear matter and $\gamma=2$ for neutron matter.
Here we have used that the mean field $\overline{\omega}$ created
by uniformly distributed nucleons is determined by baryon number 
conservation to be\cite{guichon88,fleck90,saito94}
\begin{equation}
\overline{\omega}={3\, q^q_\omega\, \rho_N\over m_\omega^2}
= {g_\omega\,\rho_N\over m_\omega^2}\ ,
\label{vec-field}
\end{equation}
where $g_\omega\equiv 3 g^q_\omega$.
The scalar mean field is determined by the thermodynamic condition
%
$\left(\partial\, E_{\rm tot}/ 
\partial\, \overline{\sigma} \right)_{R,\rho_N} = 0$,
%
which yields the self-consistency condition 
\begin{equation}
g_\sigma\, \overline{\sigma} = {g_\sigma^2\over m_\sigma^2}\, 
C(\overline{\sigma})
{\gamma\over (2\pi)^3} 
\int^{k_F}\, d^3 k\, {M_N^*\over 
\sqrt{M_N^{* 2}+{\bf k}^2}} \ ,
\label{scc}
\end{equation}
where $g_\sigma\equiv 3\, g^q_\sigma$ and 
\begin{equation}
C(\overline{\sigma}) =
{E_{\rm bag}\over M_N^*}
\Biggl[\left(1-{\Omega_q\over E_{\rm bag}\, R}\right)\,
S(\overline{\sigma}) +
{m_q^*\over E_{\rm bag}}\Biggr]
\Biggl[1-{\kappa\over 3}\, {E_{\rm bag}\over M_N^{* 2}}
{4\over 3}\,\pi\,R^3\, B\,
\Biggr]^{-1}\ ,
\label{tc}
\end{equation}
with 
\begin{equation}
S(\overline{\sigma}) = 
{\Omega_q/2+R\,m_q^*\,(\Omega_q-1)\over
\Omega_q\,(\Omega_q-1)+R\,m^*_q/2}\ .
\end{equation}

We now turn to present numerical results. For simplicity, we take
$m_q^0=0$. The coupling constants $g_\sigma$ and $g_\omega$ are
chosen to fit the nuclear matter binding energy ($-16$ MeV) at the
saturation density ($\rho_{\rm N}^0=$0.17 fm$^{-3}$). The resulting 
coupling constants and the nuclear incompressibility 
are listed in Table 1. We note that while the scalar coupling
decreases, the vector coupling and the nuclear incompressibility 
increase as $\kappa$ increases.

\begin{table}
\caption{Coupling constants and nuclear incompressibility $K$ (in MeV)
at various $\kappa$ values. The case of $\kappa=0$ corresponds to the 
simple QMC model and the last raw gives the result of QHD-I. Here we have used
$m_\sigma=550$ MeV and $m_\omega=783$ MeV.}
\label{tab-1}
\begin{tabular}{cccccccccc}
 $\kappa$ value             &\multicolumn{3}{c}{$R_0$=0.6 fm}
&\multicolumn{3}{c}{$R_0$=0.8 fm}&\multicolumn{3}{c}{$R_0$=1.0 fm}\\
              &$g_\sigma^2/4\pi$&$g_\omega^2/4\pi$ &$K$ 
&$g_\sigma^2/4\pi$&$g_\omega^2/4\pi$ &$K$ 
&$g_\sigma^2/4\pi$&$g_\omega^2/4\pi$ &$K$ \\
\tableline
$\kappa=0$     & 20.17 & 1.56 & 223 & 21.87 & 1.14 & 200 &22.47 & 0.96 & 189\\
%
%
$\kappa=7$     & 4.40  & 4.41 & 356 & 4.51  & 3.51& 314 &4.54 &3.18 & 304\\
$\kappa=8$     & 3.13  & 5.72 & 426 & 3.14  & 4.50& 359 &3.14 &4.07 & 350\\
$\kappa=9$     & 2.10  & 8.32 & 650 & 2.04  & 6.33& 471 &2.00 &5.67 & 415\\
$\kappa=10$    & 1.21  & 14.89& 2233 & 1.20 &11.17& 1058 &1.16 & 9.74 & 918\\
QHD-I          & 8.45  &12.84 &540 & 8.45  &12.84& 540 & 8.45  &12.84& 540\\
\end{tabular}
\end{table}

\begin{figure}[t]
\begin{center}
\epsfysize=11.7truecm
\leavevmode
\setbox\rotbox=\vbox{\epsfbox{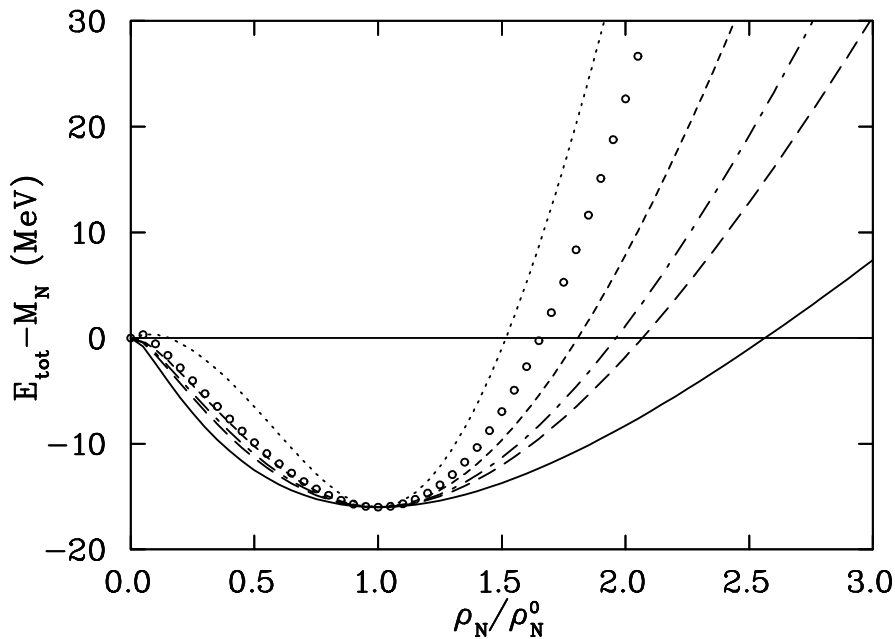}}\rotl\rotbox
\end{center}
\caption{Energy per nucleon for symmetric nuclear matter as a
function of the medium density, with $R_0=0.8$ fm. The five
curves correspond to $\kappa=0$ (solid), 7 (long-dashed), 
8 (dott-dashed), 9 (short-dashed), and 10 (dotted), respectively.
The result from QHD-I is given by the open circles.}
\label{fig-1}
\end{figure}
\begin{figure}[t]
\begin{center}
\epsfysize=11.7truecm
\leavevmode
\setbox\rotbox=\vbox{\epsfbox{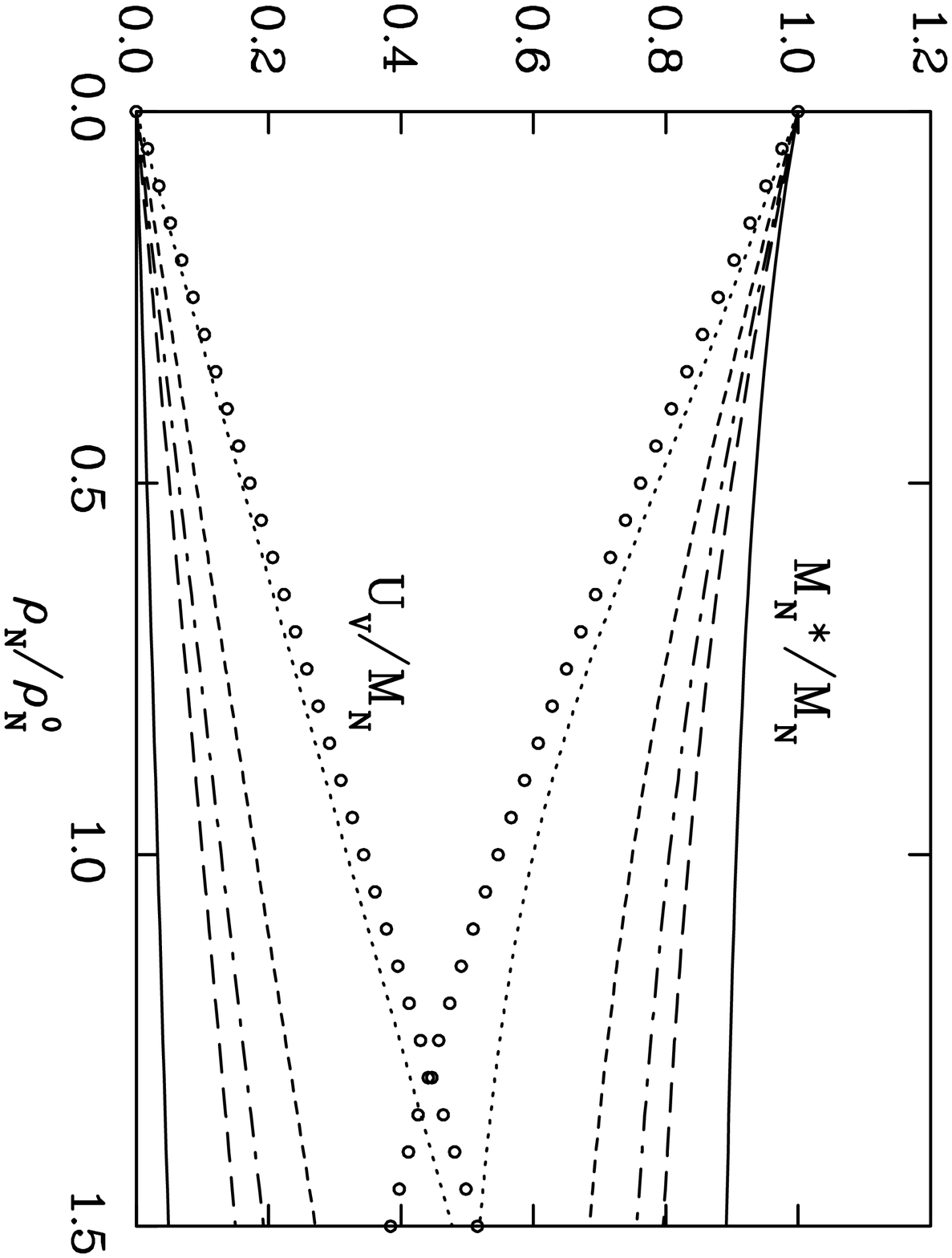}}\rotl\rotbox
\end{center}
\caption{Results for the ratios $M_N^*/M_N$ and 
$U_{\rm v}/M_N\,\equiv g_\omega\, \overline{\omega}/M_N$ as functions
of the medium density,  with $R_0=0.8$ fm. The five
curves correspond to, $\kappa=0$ (solid), 7 (long-dashed), 
8 (dot-dashed), 9 (short-dashed), and 10 (dotted), respectively.
The result from QHD-I is given by the open circles.}
\label{fig-2}
\end{figure}

This result can be understood from the scaling ansatz Eq.~(\ref{ansatz}).  When
$\kappa=0$, the strength of the vector field is much smaller than that required
in QHD-I. This, in Refs.~\cite{fleck90,saito94}, is attributed to the repulsion
provided by the c.m. corrections to the bag energy. When $\kappa > 0$,
Eq.~(\ref{ansatz}) provides a new source of attraction as it effectively
reduces $M_N^*$. Consequently, additional vector field strength is required to
balance the nuclear matter. The decrease of the scalar coupling with increasing 
$\kappa$ is due to the increasingly strong attraction from the dropping bag constant.
(When $\kappa > 10$, the self-consistent solution around $\rho_N=\rho_N^0$ does not 
exist as the attraction gets too strong.) The rapid increase of the nuclear 
incompressibility with increasing $\kappa$ results from that the contribution 
of the vector field to the nuclear incompressibility is proportional to $g_\omega^2$.

The total energy per nucleon for symmetric nuclear matter is presented in 
Fig.~\ref{fig-1} for various $\kappa$ values, with $R_0=0.8$ fm. The result 
from QHD-I is also plotted for comparison. The usual QMC model ($\kappa=0$) 
predicts a much softer equation of state for the nuclear matter than 
in QHD-I. As $\kappa$ gets larger, the equation of state  becomes stiffer, and 
when $\kappa\sim 9.5$ the equation of state is essentially the same as the one
predicted in QHD-I. The resulting
effective mass and the vector field strength are shown in Fig.~\ref{fig-2}.
One can see clearly that the effective mass decreases and the 
vector field strength increases rapidly as $\kappa$ increases.
As shown in Ref.~\cite{guichon95}, the equivalent scalar and vector 
potentials appearing in the wave equation for a point-like nucleon 
are $M_N^*-M_N$ and $U_{\rm v}\,\equiv g_\omega\,\overline{\omega}$,
respectively. Thus, our results indicate that the scalar and
vector potentials for the nucleon become stronger as $\kappa$ 
gets larger. 

\begin{figure}[t]
\begin{center}
\epsfysize=11.7truecm
\leavevmode
\setbox\rotbox=\vbox{\epsfbox{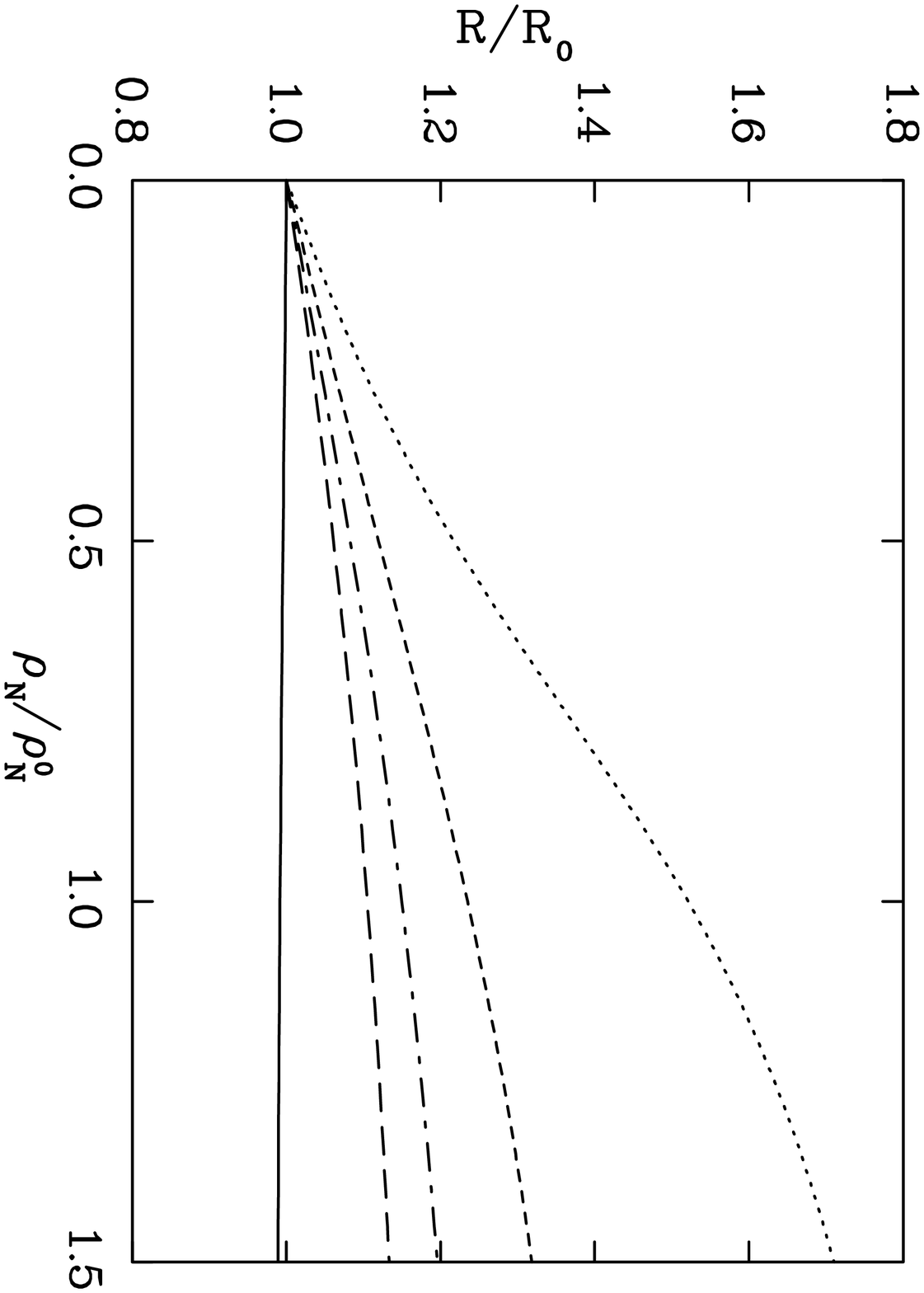}}\rotl\rotbox
\end{center}
\caption{Result for the ratio $R/R_0$ as a function of 
the medium density, with $R_0=0.8$ fm. 
The five
curves correspond to $\kappa=0$ (solid), 7 (long-dashed), 
8 (dot-dashed), 9 (short-dashed), and 10 (dotted), respectively.}
\label{fig-3}
\end{figure}
\begin{figure}[b]
\begin{center}
\epsfysize=11.7truecm
\leavevmode
\setbox\rotbox=\vbox{\epsfbox{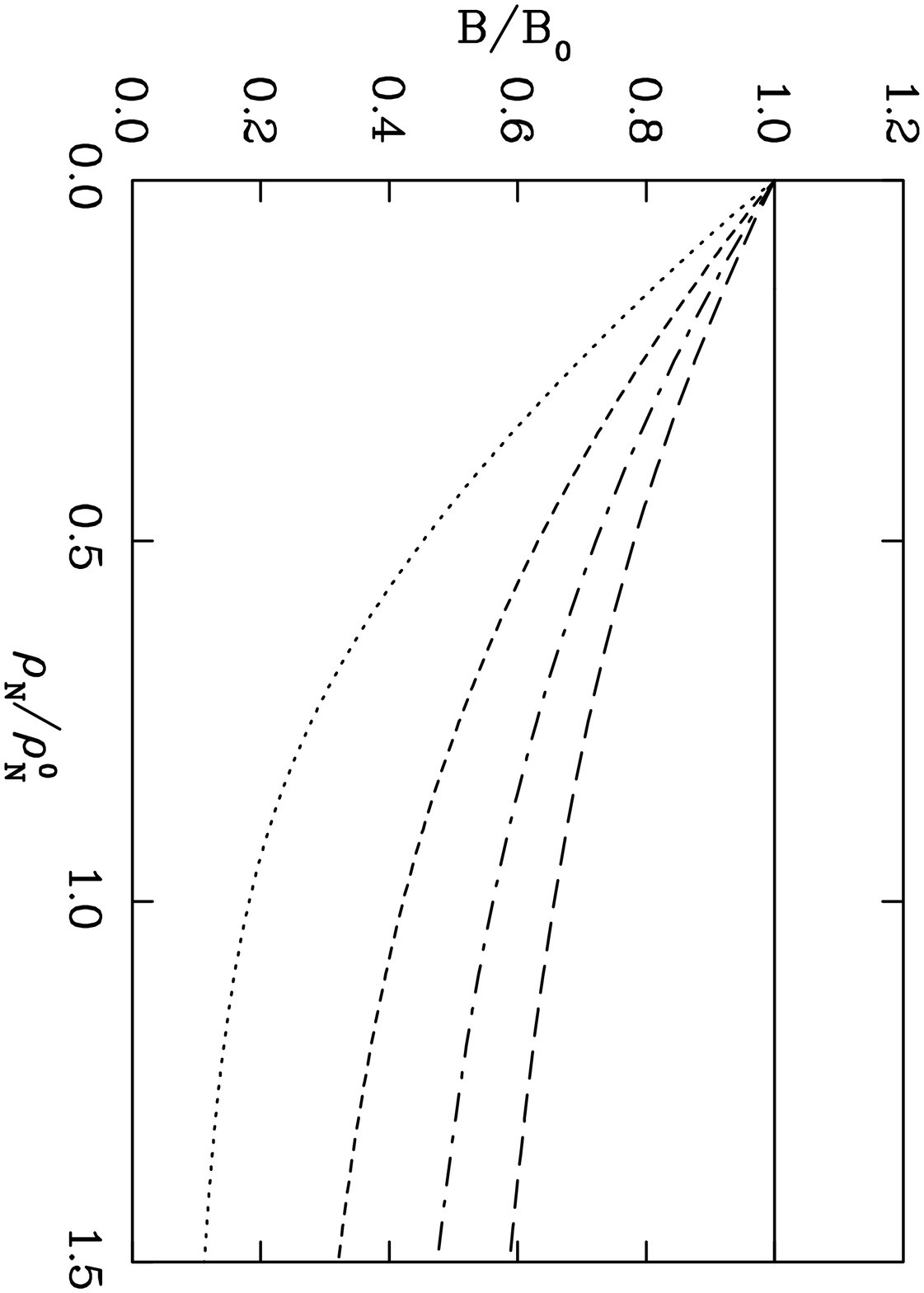}}\rotl\rotbox
\end{center}
\caption{Result for the ratio $B/B_0$ as a function of 
the medium density, with $R_0=0.8$ fm. 
The five
curves correspond to $\kappa=0$ (solid), 7 (long-dashed), 
8 (dot-dashed), 9 (short-dashed), and 10 (dotted), respectively.}
\label{fig-4}
\end{figure}

\begin{figure}[t]
\begin{center}
\epsfysize=11.7truecm
\leavevmode
\setbox\rotbox=\vbox{\epsfbox{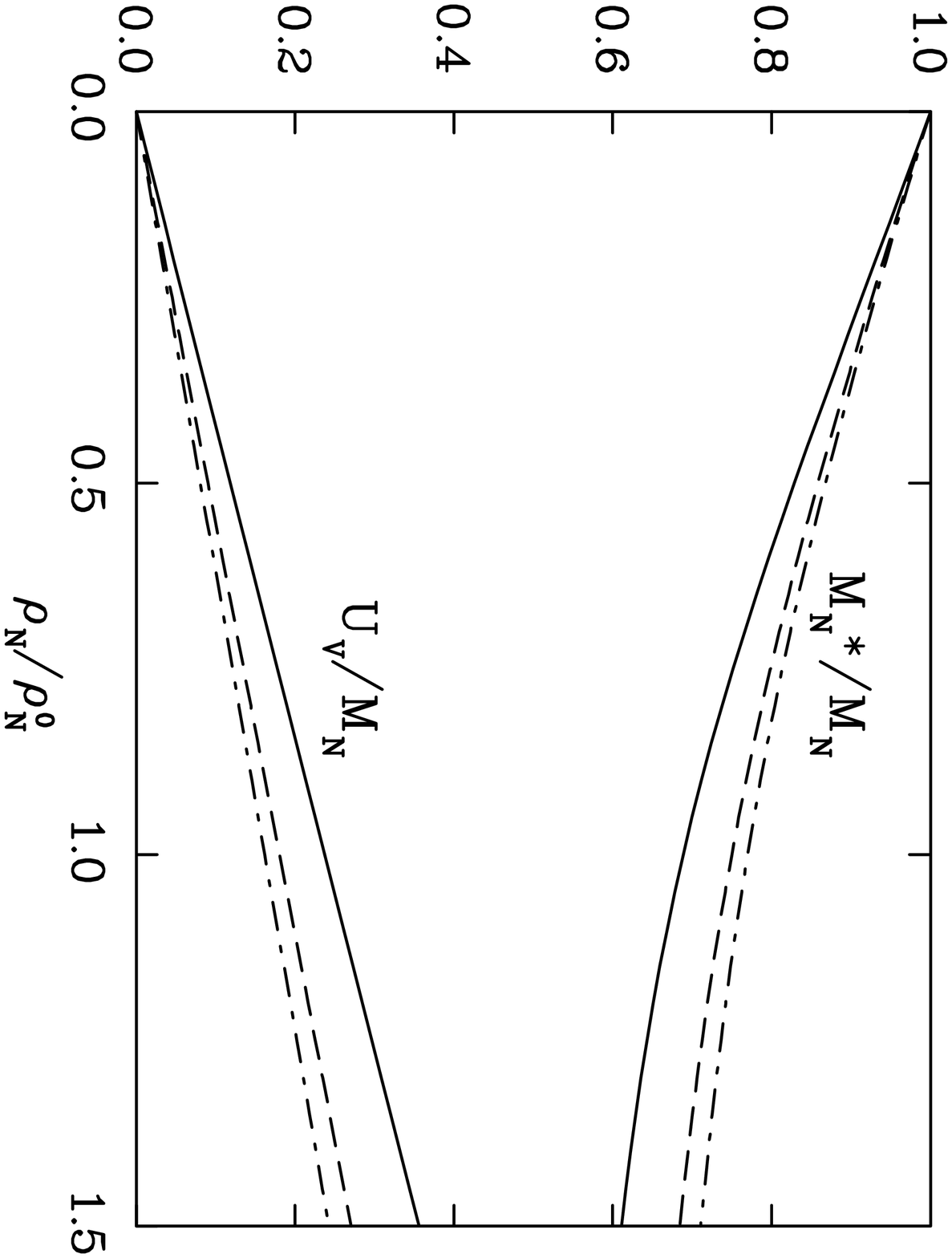}}\rotl\rotbox
\end{center}
\caption{Results for the ratios $M_N^*/M_N$ and 
$U_{\rm v}/M_N\,\equiv g_\omega\,\overline{\omega}/M_N$ as functions
of the medium density,  with $\kappa=9.0$. The three curves
are for $R_0 = 0.6$ fm (solid), 0.8 fm (dashed), and
1.0 fm (dot-dashed), respectively.}
\label{fig-5}
\end{figure}

The corresponding in-medium bag radius is shown in Fig.~\ref{fig-3}.  In the
QMC model, the bag radius decreases slightly in the medium.  When $\kappa > 0$,
the bag constant drops relative to its free space value, which implies a
decreasing bag pressure and hence gives rise to a increasing bag radius in the
medium.  With larger $\kappa$, the bag radius grows more quickly.  This is
consistent with the ``swollen'' nucleon picture drawn from the decrease of the
meson masses in nuclear medium
\cite{brown91,altemus80,reffay88,noble81,celenza85,brown89,soyeur93,brown88,%
hatsuda92,kurasawa88} (see, however, Ref.~\cite{sick85}).
The in-medium bag constant is plotted in Fig.~\ref{fig-4}. One can see that
when $\kappa > 0$ the bag constant decreases as density increases. The rate of
this decrease gets larger for larger $\kappa$ values. Finally, the sensitivity 
of our results to the free space bag radius is illustrated in Fig.~\ref{fig-5}. 
For a given $\kappa$ value, one finds that the ratios $B/B_0$ and $M_N^*/M_N$ 
increase and the ratio $R/R_0$ and the vector field strength decrease as $R_0$ 
increases.

We observe that $\kappa\sim 0$ and $\kappa\sim 7-10$ lead to qualitatively
different physics. Unless one expresses the bag constant in terms of QCD
operators and solves QCD in the nuclear matter, $\kappa$ is
unknown.  Nevertheless, one may get an estimate from the Brown-Rho scaling
ansatz, $\Phi\simeq m_\rho^*/m_\rho$. Using $m_\rho^*/m_\rho\sim 0.8$
at $\rho_N=\rho_N^0$ as suggested in Ref.~\cite{hatsuda92} and 
$B/B_0\simeq \Phi^4$, we get $B/B_0\simeq 0.4$ at the saturation density.
This requires $\kappa\simeq 8.6,\, 9.1,\, 9.3$ for $R_0= 0.6,\, 0.8,\, 1.0$ fm, 
respectively. For these $\kappa$ values, the reduction of the bag constant
dominates the attraction, the internal structure of the nucleon only plays
a relatively minor role, and the large and canceling scalar and vector
potentials for the nucleon appears naturally. Such potentials are 
comparable to those suggested by Dirac phenomenology \cite{hama90,wallace87}, 
Brueckner calculations \cite{wallace87}, and finite-density QCD sum 
rules \cite{cohen95}, but somewhat smaller than those obtained in QHD-I.
These large potentials also imply a strong nucleon spin-orbit potential.  
Therefore, we conclude that the essential features of relativistic nuclear phenomenology
can be recovered when the decrease of the bag constant with increasing density 
is considered.

The QMC model is valid only if the nucleon bags do not overlap significantly.
With the $\kappa$ values suggested above, we find $R/R_0\sim 1.25$ at
$\rho_N=\rho_N^0$. For $R_0=0.6-0.7$ fm, as suggested by Guichon \cite{guichon88},
one finds $R\sim 0.75-0.875$ fm, which gives $4 \pi R^3 \rho^0_N/3\sim 0.3-0.48$.
This indicates that the overlap between the bags is reasonably small at the saturation
density, though a factor of two larger than in the usual QMC model. For larger 
$R_0$ and/or higher densities, the overlap becomes more significant and the 
non-overlapping picture of the nuclear matter may become inadequate.
However, it is unclear at this stage whether the overlap between the bags 
is effectively included in the scalar and vector mean fields. Further study is 
needed to clarify this issue. We also note that for the $\kappa$ values 
suggested above, the resulting nuclear incompressibility is comparable to 
that obtained in QHD-I, which is too large compared with the empirical value
and that obtained in the usual QMC model. This may be fixed by introducing
self-interactions of the scalar field, which, however, will introduce more
free parameters.
 
In summary, we have included the decrease of the bag constant in the
quark-meson coupling model for the nuclear matter. This effectively introduces
a new source of attraction, which needs to be compensated with additional
vector field strength. When the change of the bag constant is large, as
supported by partial chiral-symmetry restoration, large and canceling scalar
and vector potentials for the nucleon emerge. The essential physics of the
relativistic nuclear phenomenology can be recovered.  The internal quark
structure of the nucleon seems to play only a relatively minor role. On the other hand,
the in-medium modification of the bag constant may play an important role
in low- and medium-energy nuclear physics. The model presented in the present
paper can be applied to variety of nuclear physics problems. Work along this
direction is in progress and will be reported elsewhere.

\vspace*{1cm}
This work was supported by the Natural Sciences and Engineering 
Research Council of Canada.


\begin{references}
%
\bibitem{hama90}
S. Hama, B.~C. Clark, E.~D. Cooper, H.~S. Sherif, and
     R.~L. Mercer, Phys.\ Rev.\ C 41 (1990) 2737, and references
therein.
%
\bibitem{wallace87}S.~J. Wallace, Ann.\ Rev.\ Nucl.\ Part.\ Sci.\ 
        37 (1987) 267, and references therein.
%
\bibitem{serot86}B.~D. Serot and J.~D. Walecka,
                 Adv.\ Nucl.\ Phys.\ 16 (1986) 1,
and references therein. 
%
\bibitem{cohen95}T.~D. Cohen, R.~J. Furnstahl,
D.~K. Griegel, and 
X.~Jin, Prog.\ Part.\ Nucl.\ Phys.\ Vol. 35 (1995) 221, and
references therein;
 R.~J. Furnstahl, X.~Jin, and D.~B.~Leinweber, 
nucl-th/9511007.
%
\bibitem{guichon88}P.~A.~M. Guichon, Phys. Lett. B 200 (1988) 235.
%
\bibitem{fleck90}S.~Fleck, W.~Bentz, K.~Shimizu, and K.~Yazaki,
Nucl. Phys. A 510 (1990) 731.
%
\bibitem{saito94}K.~Saito and A.~W.~Thomas, Phys. Lett. B 327 (1994) 9.
%
\bibitem{saito92}K.~Saito, A.~Michels, and A.~W.~Thomas,
Phys. Rev. C 46 (1992) R2149; 
K.~Saito and A.~W.~Thomas, Nucl. Phys. A 574 (1994) 659.
%
\bibitem{saito94a}K.~Saito and A.~W.~Thomas, Phys. Lett. B 335 (1994) 17.
%
\bibitem{saito95}K.~Saito and A.~W.~Thomas, Phys. Lett. B 363 (1995) 157.
%
\bibitem{saito95a}K.~Saito and A.~W.~Thomas,
Phys. Rev. C 51 (1995) 2757.
%
\bibitem{song95}H.~Q.~Song and R.~K.~Su, Phys. Lett. B 358 (1995) 179.
%
\bibitem{guichon95}P.~A.~M. Guichon, K.~Saito, E.~Rodionov, and 
A.~W.~Thomas, University of Adelaide preprint ADP-95-45/T194 (1995),
nucl-th/9509034.
%
\bibitem{banerjee92}M.~K.~Banerjee, Phys. Rev. C 45 (1992) 1359;
V.~K.~Mishra, Phys. Rev. C 46 (1992) 1143;
E.~Naar and M.~C.~Birse, J. Phys. G 19 (1993) 555.
%
\bibitem{adami93}C. Adami and G.~E.~Brown, Phys. Repts. 234 (1993) 1,
and references therein.
%
\bibitem{brown95}G.~E.~Brown, M.~Buballa, Z.~Li, and J.~Wambach,
Nucl. Phys. A 593 (1995) 295; G.~E.~Brown and Mannque Rho, hep-ph/9504250,
and references therein.
%
\bibitem{brown91}G.~E.~Brown and M.~Rho, Phys.\ Rev.\ Lett. 66 (1991) 2720.
%
\bibitem{altemus80} R. Altemus {\it et al.\/}, Phys.\ Rev.\ Lett.\
    44 (1980) 965; P. Barreau {\it et al.\/}, Nucl.\ Phys.\
    A 402 (1983) 515; Z.~E.  Meziani {\it et al.\/}, Phys.\
    Rev.\ Lett.\ 52 (1984) 2130; Z.~E.  Meziani {\it et al.\/},
    Phys.\ Rev.\ Lett.\ 54 (1985) 1233; A. Zghiche {\it et
    al.\/}, Nucl.\ Phys.\ A 572 (1994) 513.
%
\bibitem{reffay88} D. Reffay-Pikeroen {\it et al.}, Phys.\ Rev.\
  Lett.\ 60 (1988) 776; A. Magnon {\it et al.}, Phys.\ Lett.\
  B 222 (1989) 352; J.~E. Ducret {\it et al.}, Nucl.\ Phys.\
  A 556 (1993) 373.
%
\bibitem{noble81} J.~V. Noble, Phys.\ Rev.\ Lett.\ 46 (1981) 412.
%
\bibitem{celenza85} L.~S. Celenza, A. Rosenthal, and C.~M. Shakin,
 Phys.\ Rev.\ C 31 (1985) 232.
%
\bibitem{brown89} G.~E. Brown and M. Rho, Phys.\ Lett.\ B 222 (1989) 324.
%
\bibitem{soyeur93} M. Soyeur, G.~E. Brown, and M. Rho, Nucl.\ Phys.\
   A556 (1993) 355.
%
\bibitem{brown88} G.~E. Brown, C.~B. Dover, P.~B. Siegel, and
    W. Wiese, Phys.\ Rev.\ Lett.\ 60 (1988) 2723.
%
\bibitem{hatsuda92} T. Hatsuda and S.~H. Lee, Phys.\ Rev.\ C 46 (1992) R34;
M. Asakawa and C.~M. Ko, Nucl.\ Phys.\ A 560
  (1993) 399;M. Asakawa and C.~M. Ko, Phys.\ Rev.\ C 48 (1993) R526;
X.~Jin and D.~B.~Leinweber, Phys. Rev. C 52 (1995) 3344.
%
\bibitem{kurasawa88}H.~Kurasawa and T.~Suzuki, Nucl.\ Phys. 
A 490 (1988) 571; Prog. Theor. Phys. 84 (1990) 1030;
K.~Tanaka, W.~Bentz, A.~Arima, and F.~Beck,
Nucl. Phys. A 528 (1991) 676;
J.~C.~Caillon and J.~Labarsouque, Phys. Lett. 
B 311 (1993) 19;
H.~-C.~Jean, J.~Piekarewicz, and A.~G.~Williams,
Phys. Rev. C 49 (1994) 1981;
%
H.~Shiomi and T.~Hatsuda, Phys.\ Lett.\ B 334 (1994) 281.
%
\bibitem{sick85}I. Sick, Phys. Lett. 157B (1985) 13;
C. Atti, Nucl. Phys. A 479 (1989) 349c;
A. Zghiche {\it et al.}, Nucl. Phys. A 572 (1994) 513;
S. Ishikawa {\it et al.}, Phys. Lett. B 339 (1995) 293.

\end{references}
\end{document}